\documentclass[aps,showkeys,twocolumn,preprintnumbers,amsmath,amssymb,superscriptaddress,floatfix,nofootinbib]{revtex4-2}

\usepackage{graphicx,color,booktabs,bm}
\usepackage{amsmath}
\usepackage{slashed}
\usepackage{txfonts}
\usepackage{overpic}
\usepackage{epsfig}
\usepackage{amssymb}
\usepackage{bbding}
\usepackage{epstopdf}
\usepackage{appendix}
\usepackage{indentfirst}
\usepackage{feynmf}
\usepackage{cases}
\usepackage{multirow}

\graphicspath{{Figures/}}

\usepackage[colorlinks=true, citecolor=blue, linkcolor=red, urlcolor=blue]{hyperref}

\begin{document}
\title{Exploring interference between $P_{c \bar c}$ resonances and background in the pion-induced reaction}

\author{Qi-Fang L\"{u}}\email{lvqifang@hunnu.edu.cn}
\affiliation{Department of Physics, Hunan Normal University, Changsha 410081, China}
\affiliation{Key Laboratory of Low-Dimensional Quantum Structures and Quantum Control of Ministry of Education, Changsha 410081, China}
\affiliation{Key Laboratory for Matter Microstructure and Function of Hunan Province, and Hunan Research Center of the Basic Discipline for Quantum Effects and Quantum Technologies, Hunan Normal University, Changsha 410081, China}

\author{Xurong Chen}\email{xchen@impcas.ac.cn}
\affiliation{Southern Center for Nuclear Science Theory (SCNT),	Institute of Modern Physics, Chinese Academy of Sciences, Huizhou 516000, China}
\affiliation{State Key Laboratory of Heavy Ion Science and Technology, Institute of Modern Physics, Chinese Academy of Sciences, Lanzhou 730000, China}
\affiliation{School of Nuclear Science and Technology, University of Chinese Academy of Sciences, Beijing 100049, China}

\author{Yu-Bing Dong}\email{dongyb@ihep.ac.cn}
\affiliation{Institute of High Energy Physics, Chinese Academy of Sciences, Beijing 100049, China}
\affiliation{School of Physical Sciences, University of Chinese Academy of Sciences, Beijing 101408, China}

\author{Igor I. Strakovsky}\email{igor@gwu.edu}
\affiliation{Institute for Nuclear Studies, Department of Physics, The George Washington University, Washington, District of Columbia 20052, USA}

\author{Xiao-Yun Wang}\email{xywang@lut.edu.cn}
\affiliation{Department of Physics, Lanzhou University of Technology, Lanzhou 730050, China}
\affiliation{Lanzhou Center for Theoretical Physics, Key Laboratory of Theoretical Physics of Gansu Province, and Key Laboratory of Quantum Theory and Applications of MoE, Lanzhou University, Lanzhou 730000, China}

\author{Ju-Jun Xie}\email{xiejujun@impcas.ac.cn}
\affiliation{Southern Center for Nuclear Science Theory (SCNT),	Institute of Modern Physics, Chinese Academy of Sciences, Huizhou 516000, China}
\affiliation{State Key Laboratory of Heavy Ion Science and Technology, Institute of Modern Physics, Chinese Academy of Sciences, Lanzhou 730000, China}
\affiliation{School of Nuclear Science and Technology, University of Chinese Academy of Sciences, Beijing 100049, China}

\begin{abstract}
	
We present a phenomenological study of the near-threshold $\pi^- p \to J/\psi n$ reaction, focusing on the interference effects between the $s-$channel $P_{c \bar c}$ resonances and non-resonance background. This reaction provides a complementary probe to photoproduction for investigating the nature of the hidden-charm pentaquark states. Using an effective Lagrangian approach, we construct scattering amplitudes for $J^P=1/2^-$ and $3/2^-$ $P_{c\bar c}$ states under two spin-parity scenarios, with the background dominated by the $t-$channel $\rho$ meson exchange. We find that the interference can produce peaks, dips, or complex line shapes depending on the relative phase, in contrast to the simple resonance peaks commonly reported in previous theoretical studies. The two spin-parity scenarios yield qualitatively similar patterns, indicating that our main conclusion is robust. Given the current scarcity of experimental data for the pion-induced channel, we emphasize the urgent need for future measurements at J-PARC to understand the dynamics of the $J/\psi N$ interaction near threshold and to search for the heavy pentaquark states.

\end{abstract}

\keywords{$P_{c\bar c}$ resonances, pion-induced reaction, interference}

\maketitle

\section{introduction}

The exploration of exotic hadrons, particularly hidden-charm pentaquarks, has emerged as a frontier in understanding the non-perturbative dynamics of Quantum Chromodynamics. Since their discovery by the LHCb collaboration in the $\Lambda_b^0 \to J/\psi K^- p$ decay~\cite{LHCb:2015yax,LHCb:2019kea}, the $P_{c \bar c}$ states have challenged the conventional three-quark picture for baryons in the constituent quark model, offering a unique laboratory to study more exotic configurations composed with quarks. The proximity to various $\Sigma_c^{(*)} \bar D^{(*)}$ thresholds has fueled intense debate regarding their internal stuctures, such as weakly bound molecular states, compact multiquark states, or kinematical effects. More details can be found in review articles and previous studies; see, for instance, Refs.~\cite{Guo:2017jvc,Liu:2019zoy,Brambilla:2019esw,Guo:2019twa,Chen:2022asf,Meng:2022ozq,Liu:2024uxn,Wang:2025sic,Chen:2015moa,Chen:2015loa,Xiao:2019aya,Liu:2019tjn,Du:2019pij,Yamaguchi:2019seo,Ali:2016dkf,Wang:2015epa,Santopinto:2016pkp,Park:2017jbn,Yang:2015bmv,Wang:2019got,Liu:2015fea}. Unraveling the nature of these $P_{c \bar c}$ states is essential for a complete picture of strong interactions in the low-energy regime.

Experimentally, the evidence for $P_{c \bar c}$ states has been predominantly accumulated through heavy-flavor weak decays, which is rich in statistics but involve complex multi-body final states. Processes like the $\Lambda_b^0$ weak decays, are inherently susceptible to intricate kinematic effects, such as the troublesome triangle singularity~\cite{Guo:2019twa}, which can manifests as a spurious peak or dip in the cross sections or invariant mass distributions~\cite{Azimov:2009ta}. Consequently, extracting the genuine properties of $P_{c \bar c}$ states from these decay amplitudes often relies on model-dependent partial-wave analyses, leaving their underlying dynamics open to alternative interpretations. This ambiguity highlights the critical need for complementary production mechanisms.

A more direct way to investigate the $P_{c\bar c}$ production mechanism is through exclusive two-body scattering processes. The photoproduction reaction $\gamma p \to J/\psi p$ was therefore immediately recognized as an ideal platform. This process allows for a clean $s$-channel formation of the $P_{c\bar c}$ states, which are then observed via their decay into $J/\psi p$. Following the LHCb discovery, numerous theoretical works have investigated the $\gamma p \to J/\psi p$ reaction~\cite{Wang:2015jsa,Kubarovsky:2015aaa,Karliner:2015voa,HillerBlin:2016odx,Winney:2019edt,Wang:2019krd,Wu:2019adv,Cao:2019kst,Paryev:2022wov,Strakovsky:2023kqu,Duan:2024hby,Zhang:2024dkm,Strakovsky:2025rsm,Sakinah:2026bhv,Clymton:2026kka,Tan:2026qba,Wang:2026vuk,Strakovsky:2026qkv}. Most of these works predicted that the $P_{c \bar c}$ states would manifest as clear resonance peaks in the total cross section. However, the high-statistics measurements from the GlueX experiments presented a significant puzzle: no such resonant peak structures were observed~\cite{GlueX:2019mkq,GlueX:2023pev}. Moreover, the recent CLAS measurement on the neutron yields cross sections consistent with the proton results within uncertainties, although the limited statistics do not yet allow for a firm conclusion regarding isospin breaking effects.~\cite{CLAS:2026bis}.

Nearly all theoretical analyses of the GlueX data have searched for a resonance peak arising from constructive interference between the $s-$channel $P_{c\bar c}$ resonance and $t-$channel background. However, a recent reanalysis by Strakovsky et al.~\cite{Strakovsky:2023kqu} proposed a different interpretation that the $P_{c \bar c}(4312)$ state could still be present but masked by destructive interference with the background, producing a sharp dip structure. While the GlueX statistics have improved substantially compared to the earlier measurement, they remain insufficient to disentangle multiple resonance contributions, so only the lowest mass $P_{c \bar c}(4312)$ state was considered in this analysis. In addition, the recent  \(J/\psi\text{-}007\)~\cite{007:2026dow} and CLAS~\cite{CLAS:2026lls} measurements do not yet have sufficient statistics or energy coverage to confirm or rule out this dip. It is therefore evident that photoproduction alone leaves the nature of these structures highly ambiguous.

Given this ambiguity, an independent probe is urgently needed. The pion-induced reaction $\pi^- p \to J/\psi n$ offer a distinct and complementary perspective. As a purely hadronic process, it is free from the model dependence inherent in the Vector Meson Dominance (VMD) framework required for photoproduction. This provides a unique opportunity to test whether the interference effects suggested by the photoproduction data are a genuine physical phenomenon or merely an artifact of the specific reaction mechanism. 

It is worth noting that shortly after the discovery of the $P_{c \bar c}$ states, our group proposed the pion-induced reaction $\pi^- p \to J/\psi n$ as an alternative avenue to search for these exotic states, in particular their charge-neutral counterparts~\cite{Lu:2015fva}. This initiative was subsequently followed by several theoretical studies that further explored the pion induced reaction to search for $P_{c\bar c}$ resonances~\cite{Liu:2016dli,Lin:2017mtz,Kim:2016cxr,Wang:2019dsi,Pire:2022kwu,Clymton:2026ahm}. Additionally, prior to the discovery of the $P_{c \bar c}$ states, several earlier works had already studied the production mechanism of the $\pi^- p \to J/\psi n$ reaction~\cite{Kodaira:1979sf,Sibirtsev:1998cs,Wu:2013xma}, and explored the possible $N^*_{c\bar c}/P_{s\bar s}$ resonances in the relevant $\pi^- p \to D^- \Sigma_c^+/K^*\Sigma$ reactions~\cite{Garzon:2015zva,Wang:2024xvq}. Nevertheless, the theoretical efforts dedicated to the $\pi^- p \to J/\psi n$ process remain rather scarce compared to the photoproduction. On the experimental side, the situation is even more striking. Only two upper limit measurements exist, both dating back several decades with no exclusive $J/\psi$ production events ever observed~\cite{Jenkins:1977xb,Chiang:1986gn}.

Theoretical predictions for the total cross section in this channel span several orders of magnitude, reflecting a large uncertainty in the $\pi^- p \to J/\psi n$ coupling strength that remains poorly constrained. Interestingly, all post-discovery theoretical calculations predict clear resonant peak structures in the $\pi^- p \to J/\psi n$ cross section~\cite{Lu:2015fva,Liu:2016dli,Lin:2017mtz,Kim:2016cxr,Wang:2019dsi,Pire:2022kwu,Clymton:2026ahm}. Such pronounced peak expectations contrast sharply with the photoproduction case, where no peak but even a dip-like structure is observed. If one draws an analogy between the two reactions, this discrepancy raises the question of whether the pion-induced process would indeed exhibit peaks, or whether some similar interference induced dips could also emerge. This situation motivates a careful investigation of the interference effects in the $\pi^- p \to J/\psi n$ reaction. It is essential to explore whether the $P_{c\bar c}$ signals would manifest as peaks, dips or more complex structures in this reaction, as such knowledge would provide crucial guidance for future J-PARC~\cite{Aoki:2021cqa,Ryu:2025} and HHaS experiments~\cite{Chen:2025ppt}, and help to resolve the current ambiguities surrounding the nature of these states.

In this paper, we present a phenomenological framework to study the $P_{c\bar c}$ production in the pion-induced reaction. In Sec.~\ref{formalism}, we describe the theoretical framework for the $\pi^- p \to J/\psi n$ reaction, including the Lagrangian and amplitudes for the $s-$channel $P_{c\bar c}$ resonances as well as the background contribution. In Sec.~\ref{results}, we present our numerical results and discussions, with particular attention to the various interference patterns. Finally, a brief summary and outlook are given in Sec.~\ref{summary}.

\section{Theoretical framework}\label{formalism}

\begin{figure*}
	\centering
	\includegraphics[scale=0.5]{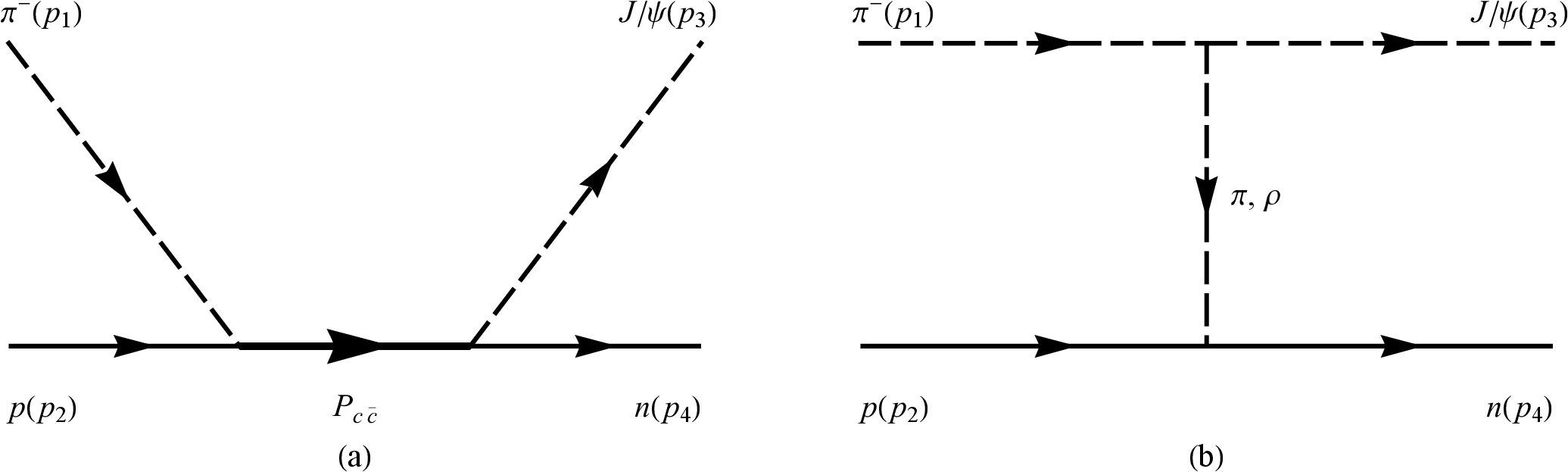}
	\caption{The relevant Feynman diagrams for the $\pi^- p \to J/\psi n$ process. The diagrams (a) and (b) correspond to the $s-$channel and $t-$channels, respectively.}
	\label{fey}
\end{figure*}

In this section, we present the theoretical formalism used to calculate the near-threshold cross section for the pion induced reaction $\pi^- p \to J/\psi n$. We adopt the effective Lagrangian approach, which provides a convenient theoretical framework to incorporate both resonances and non-resonance background contributions. The tree level Feynman diagrams considered in this work are shown in Fig.~\ref{fey}, including the $s-$channel $P_{c\bar c}$ resonances exchanges and the $t-$channel meson exchange terms. More explicitly, the signals arise from the $s-$channel three experimentally established $P_{c\bar c}$ states, $P_{c \bar c}(4312)$, $P_{c \bar c}(4440)$, and $P_{c \bar c}(4457)$, while the background contributions come from the $t-$channel $\pi$ and $\rho$ meson exchanges. 

Although the spin-parity assignments of the $P_{c \bar c}$ states have not yet been definitively determined experimentally, extensive theoretical discussions in the literature have converged on two plausible scenarios. For the three $P_{c\bar c}$ states, $P_{c \bar c}(4312)$, $P_{c \bar c}(4440)$, and $P_{c \bar c}(4457)$, the spin-parities are either $(1/2^-,1/2^-,3/2^-)$ or $(1/2^-,3/2^-,1/2^-)$. That is, the assignments for the two higher states $P_{c \bar c}(4440)$ and $P_{c \bar c}(4457)$ may be interchanged depending on the underlying theoretical interpretation. Therefore, we need the effective Lagrangians for the $P_{c\bar c}$ states with both $J^P=1/2^-$ and $3/2^-$ quantum numbers coupled to the $\pi N$ and $J/\psi N$ states. In the following, we refer to the first spin-parity assignment as Scenario A and the second as Scenario B, and we will present and compare the results for both scenarios in the next section. 

The effective Lagrangians for $P_{c \bar c} \pi N$ and  $P_{c \bar c} J/\psi N$ vertexes can be written as~\cite{Zou:2002yy,Kim:2011rm,Wang:2019dsi}:

\begin{align}
	\mathcal{L}_{P_{c\bar c}\pi N}^{1/2^-} 
	= g_{P_{c\bar c}\pi N}^{1/2^-} \bar N \vec{\tau}\cdot\vec{\pi} P_{c\bar c} + \mathrm{H.c.},
\end{align}
\begin{align}
	\mathcal{L}_{P_{c\bar c}\pi N}^{3/2^-} 
	= \frac{g_{P_{c\bar c}\pi N}^{3/2^-}}{m_\pi^2} 
	\bar N \gamma_5 \gamma_\mu \vec{\tau} \cdot \partial^\mu \partial_\nu \vec{\pi} P_{c\bar c}^\nu + \mathrm{H.c.},
\end{align}

\begin{align}
	\mathcal{L}_{P_{c\bar c}J/\psi N}^{1/2^-} = g_{P_{c\bar c}J/\psi N}^{1/2^-} \bar N \gamma_5 \gamma_\mu P_{c\bar c} \psi^\mu + \mathrm{H.c.},
\end{align}

\begin{align}
	\mathcal{L}_{P_{c\bar c}J/\psi N}^{3/2^-} = -\frac{i g_{P_{c\bar c}J/\psi N}^{3/2^-}}{2M_N} \bar N \gamma_\nu P_{c\bar c\mu} \psi^{\mu\nu} + \mathrm{H.c.},
\end{align}
with $\psi^{\mu\nu}=\partial^\mu\psi^\nu-\partial^\nu\psi^\mu$. The $\vec \tau$ is the Pauli matrix, and the $P_{c\bar c}$, $\pi$, $N$, and $\psi$ stands for the fields of $P_{c\bar c}$ resonances, pion, nucleon, and $J/\psi$ meson, respectively. For the $t-$channel $\pi$ and $\rho$ meson exchanges, the relevant Lagrangian densities can be expressed as~\cite{Liang:2004sd,Lu:2014rla,Ryu:2012tw,Wu:2013xma,Xie:2007qt}

\begin{align}
	\mathcal{L}_{J/\psi \pi \pi} 
	= -i g_{J/\psi \pi \pi} (\partial^\mu \pi^- \pi^+ - \partial^\mu \pi^+ \pi^-) \psi_\mu,
\end{align}
\begin{align}
	\mathcal{L}_{J/\psi \pi \rho} 
	= -\frac{g_{J/\psi\pi\rho}}{m_{J/\psi}} \varepsilon^{\mu\nu\alpha\beta} \partial_\mu \rho_\nu \partial_\alpha \psi_\beta \pi,
\end{align}
\begin{align}
	\mathcal{L}_{\pi NN} 
	= -\frac{g_{\pi NN}}{2M_N} \bar N \gamma_5 \gamma_\mu \vec{\tau} \cdot \partial^\mu \vec{\pi} N,
\end{align}
\begin{align}
	\mathcal{L}_{\rho NN} 
	= -g_{\rho NN} \bar N \left( \gamma_\mu + \frac{\kappa}{2M_N} \sigma_{\mu\nu} \partial^\nu \right) \vec{\tau} \cdot \vec{\rho}^\mu N,
\end{align}
where $\rho$ represents the vector field of the $\rho$ meson. 

From the above effective Lagrangian densities, the $s-$channel scattering amplitudes for $J^P=1/2^-$ and $3/2^-$ $P_{c\bar c}$ resonances can be obtained, which are written as
\begin{align}
	\mathcal{M}^{1/2^-} &= \sqrt{2} i g_{P_c\pi N} i g_{P_c J/\psi N} \, F(q^2) \, \epsilon^*_\nu(p_3,s_3) \nonumber \\
	&\quad \times \bar u(p_4,s_4) \gamma_5 \gamma_\mu S(q,M,\Gamma) u(p_2,s_2) \,,
\end{align}
\begin{align}
	\mathcal{M}^{3/2^-} &= \frac{-\sqrt{2} i g_{P_c\pi N}}{m_\pi^2} \frac{i g_{P_c J/\psi N}}{2M_N} \, F(q^2) \, \epsilon^*_\nu(p_3,s_3) \nonumber \\
	&\quad \times \bar u(p_4,s_4) \gamma_\sigma \bigl( p_3^\beta g^{\nu\sigma} - p_3^\sigma g^{\beta\nu} \bigr) \nonumber \\
	&\quad \times S_{\beta\alpha}(q,M,\Gamma) \, \gamma_5 \slashed{p}_1 p_1^\alpha u(p_2,s_2) \,,
\end{align}
with the propagators
\begin{align}
	S(q,M,\Gamma) &= \frac{i (\slashed{q} + M)}{q^2 - M^2 + i M \Gamma} \,, 
\end{align}
for the spin-1/2 fermion, and
\begin{align}
	S_{\beta\alpha}(q,M,\Gamma) &= \frac{i (\slashed{q} + M)}{q^2 - M^2 + i M \Gamma} \times \Bigg[ -g_{\beta\alpha} + \frac{1}{3}\gamma_\beta \gamma_\alpha \nonumber \\
	&\quad + \frac{1}{3M} \bigl( \gamma_\beta q_\alpha - \gamma_\alpha q_\beta \bigr) + \frac{2}{3M^2} q_\beta q_\alpha \Bigg] \,, 
\end{align}
for the spin-3/2 fermion, respectively. Here, the $p_1$, $p_2$, $p_3$, and $p_4$ are the four momenta of pion, proton, $J/\psi$, and neutron, respectively; the $s_2$, $s_3$, and $s_4$ are the spin projections of proton, $J/\psi$, and neutron, respectively; the $q=p_1+p_2$ is the four momentum of the intermediate $P_{c \bar c}$ resonances. In addition, $F(q^2)$ is the form factor reflect the finite sizes of $P_{c \bar c}$ resonances, which is commonly written as~\cite{Lu:2015fva}
\begin{align}
	F(q^2) = \frac{\Lambda_{P_{c\bar c}}^4}{\Lambda_{P_{c\bar c}}^4 + (q^2 - M^2)^2}.
\end{align}

The scattering amplitudes for $t-$channel $\pi$ and $\rho$ meson exchanges can also be expressed as 
\begin{align}
	\mathcal{M}_{\pi} 
	&= \frac{\sqrt{2} i g_{J/\psi \pi \pi} g_{\pi NN}}{M_N} 
	F_{\pi}^{NN}(q^2) F_{\pi}^{J/\psi \pi}(q^2) 
	\epsilon^*_\nu(p_3,s_3) \nonumber \\
	&\quad \times \frac{i}{q^2 - m_\pi^2} \, p_1^\nu \, \bar u(p_4,s_4) 
	\gamma_5 \slashed{q} u(p_2,s_2),
\end{align}
\begin{align}
	\mathcal{M}_{\rho} 
	&= \frac{\sqrt{2} g_{J/\psi \pi \rho} g_{\rho NN}}{m_{J/\psi}} 
	F_{\rho}^{NN}(q^2) F_{\rho}^{J/\psi \pi}(q^2) 
	\epsilon^*_\nu(p_3,s_3) \nonumber \\
	&\quad \times \varepsilon^{\alpha\beta\mu\nu} q_\alpha p_{3\mu} 
	\frac{i \left( -g_{\beta\lambda} + q_\beta q_\lambda / m_\rho^2 \right)}{q^2 - m_\rho^2} 
	\bar u(p_4,s_4) \nonumber \\
	&\quad \times \left[ \gamma^\lambda + \frac{\kappa}{4M_N} 
	\bigl( \gamma^\lambda \slashed{q} - \slashed{q} \gamma^\lambda \bigr) \right] u(p_2,s_2),
\end{align}
where $q = p_1 - p_3$ is the four-momentum transfer in the $t-$channel. The relevant form factors are 
\begin{align}
	F_m^{J/\psi \pi}(q^2) = \frac{\Lambda_m^{*2} -m^2}{\Lambda_m^{*2} -q^2}.
\end{align}
\begin{align}
	F_m^{NN}(q^2) =\Bigg(\frac{\Lambda_m^2 -m^2}{\Lambda_m^2 -q^2}\Bigg)^n.
\end{align}
with $n=1$ for pseudoscalar meson and $n=2$ for vector meson~\cite {Xie:2007qt,Gao:2026hjv}.

Then, the total scattering amplitude $\mathcal{M}_{\rm tot}$ for the $\pi^-p \to J/\psi n$ reaction is constructed as the coherent sum of the $t-$channel background and $s-$ channel resonance contributions
\begin{align}
		\mathcal{M}_{\rm tot} = \mathcal{M}_{\rm bg}+e^{i\phi}\mathcal{M}_{res},\label{amp}
\end{align}
with 
\begin{align}
	\mathcal{M}_{\rm bg} = \mathcal{M}_{\pi}+\mathcal{M}_{\rho},
\end{align}
for the background contributions, and 
\begin{align}
	\mathcal{M}_{\rm res} = \mathcal{M}_{P_{c\bar c}(4312)}+\mathcal{M}_{P_{c\bar c}(4440)}+\mathcal{M}_{P_{c\bar c}(4457)},
\end{align}
for the resonance contributions. In Eq.~(\ref{amp}), $\phi$ is a relative phase between the background and resonance amplitudes. The value of $\phi$ cannot be predicted by the present model and will be treated as a parameter to be varied in the phenomenological analysis or constrained by comparison with the future experimental data.

The differential cross section for the $\pi^-p \to J/\psi n$ reaction is then given by
\begin{align}
	\frac{d\sigma}{d\cos\theta} 
	= \frac{1}{32\pi s} 
	\frac{|\vec{p}_3^{\,\mathrm{c.m.}}|}{|\vec{p}_1^{\,\mathrm{c.m.}}|}
	\left( \frac{1}{2} \sum_{s_2,s_3,s_4} 
	|\mathcal{M}_{\rm tot}|^2 \right),
\end{align}
where $\theta$ is the scattering angle of outgoing $J/\psi$ relative to the incoming pion beam, $s$ is the Lorentz invariant Mandelstam variable defined as $s=(p_1+p_2)^2$, and $|\vec{p}_1^{\,\mathrm{c.m.}}|$ and $|\vec{p}_3^{\,\mathrm{c.m.}}|$ are the three momenta of $\pi$ and $J/\psi$ mesons in the center-of-mass frame, respectively. Finally, integrating the differential cross section over ${\rm cos} \theta$ yields the total cross section for the $\pi^- p \to J/\psi n$ process.

\section{Results and discussions}\label{results}  

In this section, we present our numerical results for the $\pi^- p \to J/\psi n$ reactions near threshold. We begin by specifying the values of the coupling constants and cutoff parameters adopted in our calculation, followed by a detailed analysis of the interference patterns between the background and $P_{c \bar c}$ resonance contributions. We then compare our findings with the photoproduction case and discuss the implications for future experiments.

\begin{figure}
	\centering
	\includegraphics[scale=0.85]{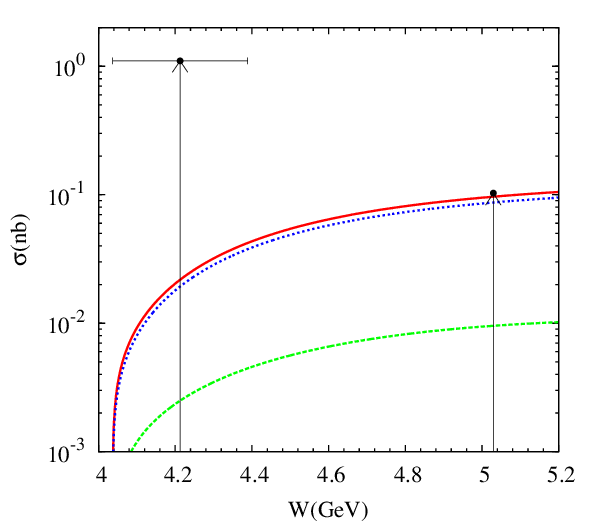}
	\caption{The background contributions for the $\pi^- p \to J/\psi n$ process together with two experimental upper limits~\cite{Jenkins:1977xb,Chiang:1986gn}. The red solid, green dashed, and blue dotted lines represent the total background, $\pi$ exchange, and $\rho$ exchange contributions, respectively.}
	\label{bac}
\end{figure}

The coupling constants and cutoff parameters used in our calculation are taken from the experimental data and literature. For the background contributions, the coupling constants of $g_{J/\psi \pi \rho}=0.032$ and $g_{J/\psi \pi \pi}=8.20\times 10^{-4}$ can be obtained from the relevant partial decay widths; the $g_{\pi NN}=13.45$, $g_{\rho NN}^2/(4\pi) = 0.9$ and $\kappa=6.1$ are commonly adopted according to Refs. \cite{Wu:2013xma,Xie:2007qt}. The cutoff parameters for the background amplitudes are constrained by the available experimental upper limits on the total cross section at $W=5.03 ~\rm{GeV}$ and near threshold region~\cite{Jenkins:1977xb,Chiang:1986gn}, where $W\equiv \sqrt{s}$ is the total center-of-mass energy of the $\pi^- p$ system. Then, we find that $\Lambda_{\rho}^* = \Lambda_{\pi}^* =1.00 $ GeV, $\Lambda_{\rho} = 1.25 $ GeV, $\Lambda_{\pi} =1.00$ GeV are consistent with the existing experimental upper limits, which is displayed in Fig.~\ref{bac}. This provides a reasonable bound on the magnitude of the non-resonant background, ensuring that our parameter choices are consistent with existing data. In addition, the $\rho$ meson exchange dominates over the $\pi$ exchange channel, consistent with the findings of previous studies. 

Since the absolute couplings of the three $P_{c\bar c}$ resonances to the $\pi N$ and $J/\psi N$ states are unknown experimentally, we assume a benchmark scenario with
\begin{align}
	{\rm Br}[P_{c \bar c} \to J/\psi N] = 10^{-2},
\end{align}
and 
\begin{align}
	{\rm Br}[P_{c \bar c} \to \pi N] = 10^{-6}.
\end{align}
Here, our assumed branching ratios of ${\rm Br}[P_{c \bar c} \to J/\psi N] $ are consistent with the model-dependent upper limit of about $2\%$ extracted from GlueX experimental data~\cite{GlueX:2019mkq}, and are also supported by theoretical calculations that typically yield branching fractions ranging from $0.1\%$ to several tens of percent~\cite{Lu:2016nnt,Xiao:2019mvs,Lin:2019qiv,Cao:2019kst,Wu:2019rog,Wang:2019spc,Dong:2020nwk,Yang:2024nss,Liang:2026yma}. To justify our choice of branching ratio of ${\rm Br}[P_{c \bar c} \to \pi N]$ , we note that the decay $P_{c \bar c} \to \pi N$ is dynamically suppressed compared to $P_{c \bar c} \to J/\psi N$. The $J/\psi N$ decay mode can proceed via a simple rearrangement of the quark constituents, while the $\pi N$ channel requires both the annihilation of the $c \bar c$ pair and the creation of light $q \bar q$ pair. These annihilation and creation processes in the $\pi N$ decay mode need occur simultaneously, which inevitably reduces the decay probability. Moreover, in the hadronic molecule scenario, the $c \bar c$ quarks are spatially separated, making the direct annihilation process even less likely. It is therefore reasonable to expect ${\rm Br}[P_{c \bar c} \to \pi N]$ to be several orders of magnitude smaller than ${\rm Br}[P_{c \bar c} \to J/\psi N]$. In addition, we adopt $10^{-6}$ as a benchmark value, for which the resonance and background amplitudes are of comparable size, thereby enabling pronounced interference effects. This corresponds to the scenario where the interference pattern is most sensitive to the relative phase and thus most amenable to experimental observation. The obtained coupling constants are listed in Table~\ref{tab1}. We will also consider alternative branching ratios with values of $10^{-5}$ and $10^{-7}$ for comparison. The cutoff parameter $\Lambda_{P_{c \bar c}}=0.5~\rm{GeV}$ is adopted as in previous works~\cite{Lu:2015fva,Wang:2019dsi}. 

\begin{table}
	\centering
	\caption{Coupling constants $|g_{P_{c \bar c}J/\psi N}|$ and $|g_{P_{c \bar c}\pi N}|$ for $P_{c \bar c}(4312)$, $P_{c \bar c}(4440)$, and $P_{c \bar c}(4457)$ resonances by assuming their decay branching ratios are $10^{-2}$ and $10^{-6}$, respectively. \label{tab1}}
\begin{tabular}{p{2.0cm} p{2.0cm} p{2.0cm} p{2.0cm}}
		\hline\hline
	Resonance       & Channel    & Scenario A & Scenario B \\
		\hline
		$P_c(4312)$ & $J/\psi N$ & $3.49\times10^{-2}$      &  $3.49\times10^{-2}$       \\
		& $\pi N$    & $1.64\times10^{-4}$ & $1.64\times10^{-4}$ \\
		$P_c(4440)$ & $J/\psi N$ &  $4.46\times10^{-2}$      & $4.63\times10^{-2}$       \\
		& $\pi N$    & $2.36\times10^{-4}$ & $1.05\times10^{-6}$ \\
		$P_c(4457)$ & $J/\psi N$ &$2.54\times10^{-2}$      &  $2.45\times10^{-2}$       \\
		& $\pi N$    & $5.78\times10^{-7}$ & $1.31\times10^{-4}$ \\
		\hline\hline
\end{tabular}
\end{table}

\begin{figure*}
	\centering
	\includegraphics[scale=0.85]{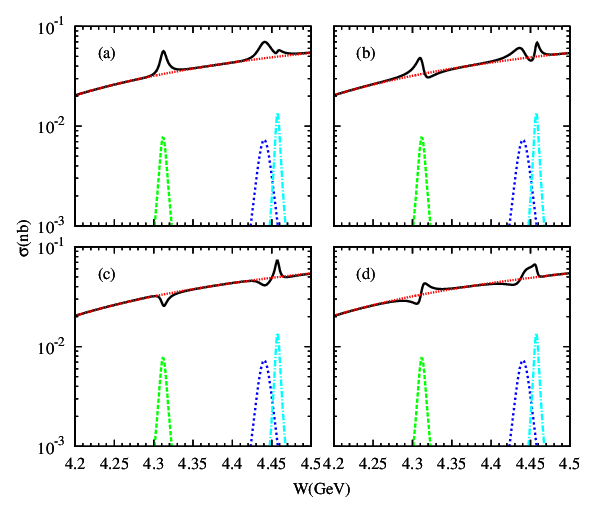}
	\includegraphics[scale=0.85]{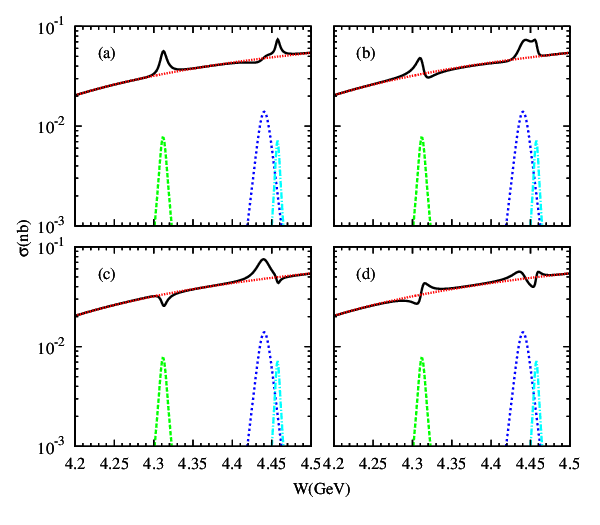}
	\caption{The cross sections for $\pi^- p \to J/\psi n$ process with significant interference. The left and right panels correspond to Scenario A and Scenario B, respectively. The (a), (b), (c), and (d) stand for the $0^\circ$, $90^\circ$, $180^\circ$, and $270^\circ$ cases, respectively. The black solid, green dashed, blue dotted, cyan dash-dotted, and red fine-dotted lines represent the total cross section, $P_{c\bar c}(4312)$,  $P_{c\bar c}(4440)$, $P_{c\bar c}(4457)$, and background contributions, respectively.}
	\label{totalab}
\end{figure*}

To illustrate the interference effects, we scan the relative phase $\phi$ between the background and resonance amplitudes over the full range from $0^\circ$ to $360^\circ$. We also restrict our analysis to the center-of-mass energy range from the threshold up to 4.5 GeV, since this interval covers the resonance region and is most suitable for demonstrating the interference patterns between the background and the 
$P_{c \bar c}$ resonance amplitudes. The results are shown in Fig.~\ref{totalab}. It can be seen that the total cross section exhibits a rich variety of structures as $\phi$ varies. The constructive interference produces pronounced peaks, destructive interference leads to dips, and intermediate phases generate asymmetric or more complex line shapes. This demonstrates that the interference pattern is highly sensitive to the relative phase, and that peaks, dips, and more complex structures can emerge depending on the relative phase $\phi$. 

The distinction between Scenarios A and B in the total cross sections around $W=4.5~\rm{GeV}$ is a direct consequence of the reversed spin-parity assignments for the $P_{c\bar c}(4440)$ and $P_{c\bar c}(4457)$ resonances. Nevertheless, both scenarios exhibit similarly complex interference structures, despite their different spin-parity assignments for the higher-mass states. This observation indicates that the qualitative conclusion of our analysis is robust. Namely, the interference between the background and resonance amplitudes can significantly distort the line shape, leading to either peaks or dips depending on the relative phase. The uncertainty in the spin-parity assignments therefore does not affect this main finding.

\begin{figure}
	\centering
	\includegraphics[scale=0.85]{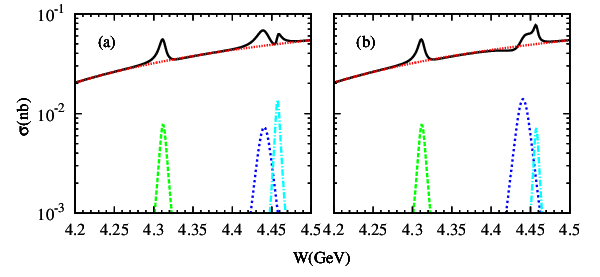}
	\caption{The cross sections for $\pi^- p \to J/\psi n$ process with relative phase $\phi=40.8^\circ$. The (a) and (b) correspond to Scenario A and Scenario B, respectively.}
	\label{total40}
\end{figure}

It is instructive to compare our results with the photoproduction reaction $\gamma p \to J/\psi p$, where a dip-like structure has been observed and interpreted as possible evidence for destructive interference. To provide an experimentally motivated benchmark, we show in Fig.~\ref{total40} the $\pi^-p \to J/\psi n$ total cross section calculated with the phase $\phi=40.8^\circ$, which was extracted from the GlueX data analysis in Ref.~\cite{Strakovsky:2023kqu} for the $P_{c\bar c}(4312)$ state. This allows a direct comparison of the line shapes in the two reactions under a specific, data-driven phase. From this comparison, we can see that the line shapes in the two reactions are not the same even when the same phase is used. This difference is not surprising, as the underlying mechanisms of the two reactions are distinct. Photoproduction proceeds via a point-like photon probe, while the pion-induced process involves a composite projectile. Nevertheless, both reactions can in principle produce either peak or dip structures, depending on the relative phase between the resonance and background amplitudes. This suggests that the absence of a peak in photoproduction does not preclude the possibility of observing a peak or a dip in the pion-induced channel. A combined analysis of both reactions could therefore provide valuable cross-checks on the interference scenario.

Note that, in our numerical calculation, all couplings involving the $P_{c \bar c}$ resonances are taken to be real and positive. This is not a restrictive assumption, as the decay width of a resonance only determines the modulus of its coupling constants as shown in Table~\ref{tab1}. Each coupling could in principle carry an additional phase factor, which is not constrained by the decay width alone. However, under the molecular interpretation of the $P_{c \bar c}$ resonances, heavy quark symmetry implies that the couplings of these resonances to a given channel are closely correlated. It is therefore reasonable to adopt a uniform sign for all $P_{c \bar c}$ couplings. Moreover, an overall common phase factor applied to all resonance couplings can be absorbed into the relative phase $\phi$ between the resonance and background amplitudes.

\begin{figure*}
	\centering
	\includegraphics[scale=0.85]{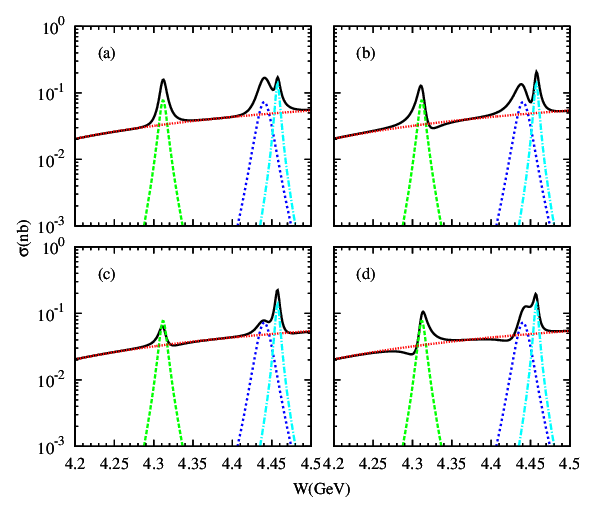}
	\includegraphics[scale=0.85]{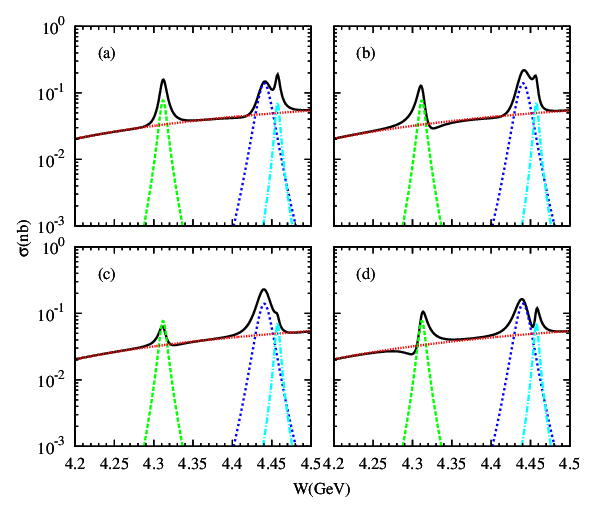}
	\caption{The notation is the same as in Fig.~\ref{totalab}, but with the ${\rm Br}[P_{c \bar c} \to \pi N]$ taken to be $10^{-5}$.}
	\label{totalcd}
\end{figure*}

\begin{figure*}
	\centering
	\includegraphics[scale=0.85]{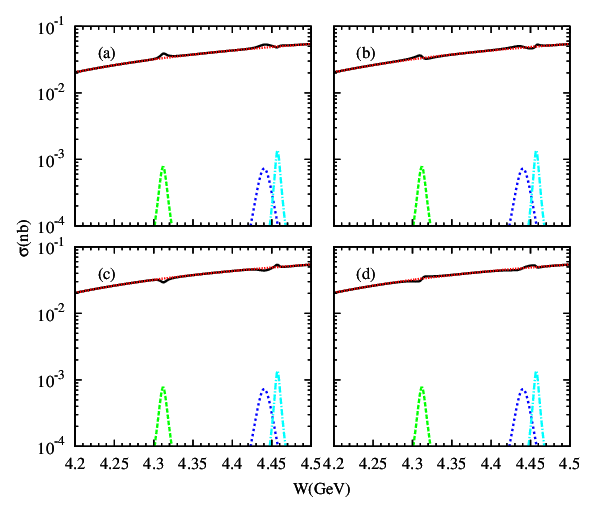}
	\includegraphics[scale=0.85]{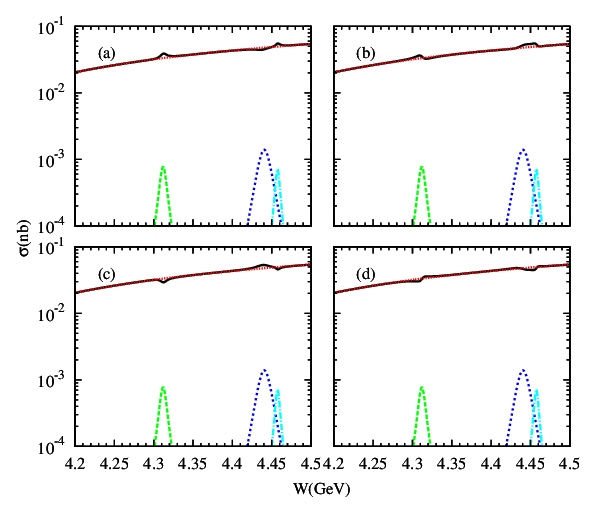}
	\caption{The notation is the same as in Fig.~\ref{totalab}, but with the ${\rm Br}[P_{c \bar c} \to \pi N]$ taken to be $10^{-7}$.}
	\label{totalef}
\end{figure*}

To investigate the impact of the resonance coupling strength on the total cross section, we artificially vary the overall magnitude of the resonance contributions. In practice, the contribution of each $P_{c \bar c}$ state scales with the product of its branching fractions to the $J/\psi N$ and $\pi N$ channels, so it is sufficient to adjust a single common factor. Here, we vary 	${\rm Br}[P_{c \bar c} \to \pi N]$ from $10^{-6}$ to $10^{-5}$ or $10^{-7}$, while keeping  ${\rm Br}[P_{c \bar c} \to J/\psi N]$ fixed at $10^{-2}$. The results are shown in Figs.~\ref{totalcd} and \ref{totalef}. When the resonance contributions dominate, the cross section exhibits clear peaks, as commonly reported in previous studies. When the resonance contributions are much smaller than the background, the signals become practically indistinguishable, making the search for $P_{c \bar c}$ states in this channel extremely challenging.

At present, experimental data for the $\pi^- p \to J/\psi n$ reaction are still very limited. With only loose constraints available, the background can be estimated roughly, while the resonance contributions and their interference patterns remain largely unconstrained. In contrast, the photoproduction process benefits from measurements by the GlueX, CLAS, and \(J/\psi\text{-}007\) Collaborations, which could provide strong constraints on theoretical models. For the pion-induced reaction, even a few data points across the near-threshold region would place powerful constraints on the model parameters, particularly on the resonance couplings and the interference phase. We strongly encourage future experiments, such as J-PARC, to measure this process. Such measurements would be crucial for clarifying the role of interference effects and the nature of the $P_{c \bar c}$ resonances.

\section{Summary}\label{summary}

In this work, we have performed a comprehensive phenomenological study of the near-threshold $\pi^- p \to J/\psi n$ reaction, with particular emphasis on the interference effects between the $s-$channel $P_{c\bar c}$ resonances and $t-$channel background. We adopted the effective Lagrangian approach and constructed the relevant amplitudes by considering two plausible spin-parity scenarios for the $P_{c \bar c}(4312)$,  $P_{c \bar c}(4440)$, and $P_{c \bar c}(4457)$ states. The background is dominated by $\rho$ meson exchange, with its parameters constrained by the available experimental upper limits. By introducing a relative phase $\phi$ between the background and resonance scattering amplitudes, we systematically investigated the dependence of the total cross section on the interference pattern.

Our analysis reveals several key findings. The interference between the non-resonant background and $P_{c \bar c}$ resonances can produce a rich variety of structures in the $\pi^- p \to J/\psi n$ cross section, including peaks, dips, and more complex line shapes, depending on the relative phase. This is in sharp contrast to the naive expectation of resonance peaks that has been commonly reported in previous theoretical studies. We also find that the two spin-parity scenarios yield quantitatively different results for the higher-mass states, but both exhibit similarly complex interference patterns, indicating that our qualitative conclusion is robust against the uncertainty in the spin-parity assignments. Compared to the photoproduction case, where a dip-like structure has been observed and interpreted as possible evidence for destructive interference, the pion-induced reaction offers a complementary probe with a different initial-state configuration and interference phase. Importantly, we have shown that the interference effect is most pronounced when the resonance and background contributions are comparable in magnitude.

Looking forward, our results underscore the urgent need for experimental data on the 
$\pi^- p \to J/\psi n$ reaction. The upcoming experiments with the high-intensity pion beam at J-PARC and HHaS promise to provide the high-statistics measurement in this channel. Even a few data points across the near-threshold region would place strong constraints on the model parameters and help clarify whether the $P_{c \bar c}$ signals manifest as peaks or dips, or exhibit more complex interference structures. On the theoretical side, the framework developed in this work can be extended to other two-body scattering processes. In particular, the interference effects could also be studied in reactions involving strange quarks, such as $K^-p \to J/\psi \Lambda$ reaction, which is not accessible in two-body photoproduction. This reaction provides a unique opportunity to investigate the strange partner of the $P_{c \bar c}$ discussed in present work, namely the $P_{c\bar c s}(4459)$, and would offer a new dimension for probing the nature of these hidden-charm pentaquark states.
 
\subsection*{ACKNOWLEDGMENTS}

This work was partly supported by the National Key R\&D Program of China (Grant No. 2023YFA1606703) and the National Natural Science Foundation of China (Grant Nos. 12575094, 12435007, 12361141819, and 12375142). This work is supported by the Scientific Research Foundation of Hunan Provincial Education Department under Grant No. 24B0063, and the Youth Talent Support Program of Hunan Normal University under Grant No. 2024QNTJ14. This work was also supported in part by the U.S. Department of Energy, Office of Science, Office of Nuclear Physics, under Award No. DE–SC0016583.

\end{document}